\documentclass[twoside,english]{paper}
\usepackage[T1]{fontenc}
\usepackage[latin9]{inputenc}
\usepackage{geometry}
\usepackage{units}
\usepackage{textcomp}
\usepackage{amstext}
\usepackage{graphicx}

\makeatletter
\newcommand{\lyxaddress}[1]{
\par {\raggedright #1
\vspace{1.4em}
\noindent\par}
}

\usepackage[numbers, sort&compress]{natbib}

\makeatother

\usepackage{babel}
\begin{document}

\title{A simple method for characterization of the magnetic field in an
ion trap using Be$^{+}$ ions}

\author{Jianwei Shen, Andrii Borodin{*}, and Stephan Schiller{*}}

\maketitle

\lyxaddress{Institut fuer Experimentalphysik, Heinrich-Heine-Universitaet Duesseldorf,
40225 Duesseldorf, Germany}

$^{*}$Correspondence addresses: andrii.borodin@uni-duesseldorf.de, step.schiller@uni-duesseldorf.de
\begin{abstract}
We demonstrate a simple method for the determination of the magnetic
field in an ion trap using laser-cooled $^{9}$Be$^{+}$ ions. The
method is not based on magnetic resonance and thus does not require
delivering radiofrequency (RF) radiation to the trap. Instead, stimulated
Raman spectroscopy is used, and only an easily generated optical sideband
of the laser cooling wave is required. The d.c. magnetic vector, averaged
over the $^{9}$Be$^{+}$ ion ensemble, is determined. Furthermore,
the field strength can be minimized and an upper limit for the field
gradient can be determined. The resolution of the method is 0.04~G
at present. The relevance for precision rovibrational spectroscopy
of molecular hydrogen ions is briefly discussed.
\end{abstract}

\section{Introduction}

Methods of magnetic field measurement are of high importance in different
fields of fundamental and applied research - for example in quantum
optics, quantum information processing, and high-resolution spectroscopic
measurements. The large variety of magnetic field sensors, utilizing
e.g. microelectromechanical systems (MEMS), the Hall effect, the giant
magnetoresistance effect, superconducting quantum interference (SQUID),
have different sensitivities and magnetic field ranges. A common natural
disadvantage of these sensors is the need to introduce them into the
region where the magnetic field is to be determined. Although many
sensors are compact, this requirement becomes critical in the case
of, e.g., experiments with trapped ions or atoms, where the presence
of vacuum complicates sensor deployment and the sensor can not be
used in situ while the experiment is operating. In this case, the
trapped ions or atoms themselves can serve as sensors, because the
magnetic field affects their level energies, and the measurement of
transition frequencies (shifts) can be used to deduce the magnetic
field. For an overview of different methods, see e.g. Ref.~\cite{WebsterEtAl2002,SchneiderEtAl2005,SchneiderThesis}.
Typically, for atoms in the linear Zeeman shift regime, the line splitting
is on the order of the Larmor precession frequency, which is 2.8~MHz/G.
If a sufficiently large magnetic field can be applied, a RF transition
can be used to probe the splitting, see e.g. Ref.~\cite{NakamuraBe+RF}.

The measurement of a small magnetic field, when the concomitant Zeeman
splitting is reduced to on the order of 10~kHz or less, is possible
in case of non-zero nuclear magnetic moment and non-zero electronic
angular momentum ($J$), because then RF radiation can be used to
measure the splitting between hyperfine states. As an example, the
hyperfine transition frequency $f_{0}$ in the ground $^{2}S_{\nicefrac{1}{2}}$
level of $^{9}$Be$^{+}$ is 1.25~GHz in zero field. Using RF spectroscopy
between hyperfine states, the frequencies can be measured with kHz
resolution or better. Since the Zeeman shifts of the hyperfine states
are on the order of MHz per Gauss, then a resolution at milli-Gauss
level is in principle possible.

Our approach for determining Zeeman shifts of trapped $^{9}$Be$^{+}$
ions is based on using stimulated Raman transitions. The $^{9}$Be$^{+}$
ion is a key element of many quantum logic, spectroscopic, and quantum
simulation studies \cite{ChouAlClock2010,ShigaItanoBollingerBe+2011,NakamuraBe+RF,BiercukEtAlQuantMemory2009}.
The method utilizes only the cooling light and a modulation sideband
and does not require delivering a RF field to the trap center, thus
simplifying the overall scheme. In fact, no additional hardware beyond
the one used for our experiments on sympathetic cooling was employed.

\section{Motivation}

In our experimental setup, $^{9}$Be$^{+}$ ions trapped in a linear
ion trap are used as coolant ions for sympathetic cooling of co-trapped
HD$^{+}$ molecular ions. HD$^{+}$ is a model system in precision
spectroscopy of cold trapped molecules for the purpose of testing
quantum theory and the time-independence of fundamental constants.
Precise measurements of transition frequencies of HD$^{+}$ require
information about the magnetic field experienced by the ions in the
trap. Therefore, it is natural to consider the coolant $^{9}$Be$^{+}$
ions as an in-situ magnetic field sensor.

For instance, in pure rotational spectroscopy in HD$^{+}$ at 1.3~THz
(see Ref.~\cite{Shen2012b}) it is of interest to resolve the Zeeman
structure of individual hyperfine lines. As a concrete example we
consider the hyperfine transition $(v=0,\, N=0,\, F=0,\: S=1,\: J=1)\rightarrow(v'=0,\, N'=1,\, F'=0,\: S'=1,\: J'=2)$
at a detuning $f-f_{0,theor}=-2.1$~MHz from the spinless rotational
transition frequency $f_{0,theor}$ (see Fig.~3 of Ref.~\cite{Shen2012b}).
It is particularly attractive, since it exhibits a small Zeeman shift
and a small Zeeman splitting. The $J_{z}=0\rightarrow J'_{z}=0$ component
of this particular transition has a quadratic Zeeman shift that is
approx. 2.9~kHz in a field of 1~G. Other metrological properties
have been discussed in \cite{BakalovSchillerQuadrupoleShift,SchillerBakalovKorobovPRL2014}.
Thus, assuming that one can set a magnetic field with the small value
$B<0.05$~G in the trap, and that the ions are in the Lamb-Dicke
regime, a shift and transition linewidth (due to a possible magnetic
field gradient) of less than 10~Hz should be achievable. At this
resolution level, a precise measurement of the hyperfine energy contribution
($-2.1$~MHz) and a test of the QED corrections including those of
order $\alpha^{6}$ relative to the nonrelativistic transition frequency
would become possible (see Ref.~\cite{KorobovHilicoKarr2014}). Note
that the closest two Zeeman components arise from transitions $J_{z}=\pm1\rightarrow$
$J'_{z}=\pm2$, with nearly opposite linear Zeeman shifts of approx.
$\mp40$~kHz/G, respectively, and would be clearly removed from the
$J_{z}=0\rightarrow J'_{z}=0$ transition \cite{Zee2BakalovEtAl2012}.
This example shows the utility of reducing the strength of the magnetic
field in an ion trap for a particular application.

\section{Experimental scheme and setup}

\begin{figure}
\begin{centering}
\includegraphics[width=8cm]{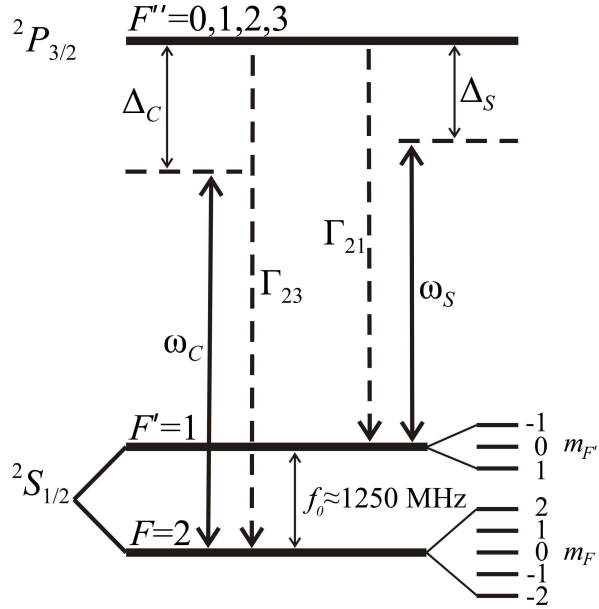}
\par\end{centering}

\protect\caption{The optical transitions relevant to the stimulated Raman spectroscopy.
$\omega_{C}$: carrier frequency; $\omega_{S}$: a red-detuned sideband
produced e.g. by an amplitude modulator or phase modulator. Only the
Zeeman splittings in the electronic ground state are shown, since
those in the electronic excited state are not relevant.}

\label{Flo:BeGeneral}
\end{figure}

The optical transitions in $^{9}$Be$^{+}$ relevant to laser cooling
are shown in Fig.~\ref{Flo:BeGeneral}. The use of $\sigma^{+}$
or $\sigma^{-}$ transitions provides an almost closed-cycle level
scheme, allowing efficient laser cooling of $^{9}$Be$^{+}$ ions.
In this case, the atomic population is almost completely distributed
between the excited $^{2}P_{\nicefrac{3}{2}}$, $F''=3$, $m_{F''}=\pm3$
and the ground $^{2}S_{\nicefrac{1}{2}}$, $F=2$, $m_{F}=\pm2$ state;
decay into $^{2}S_{\nicefrac{1}{2}}$, $F'=1$ state is forbidden,
as this would require transitions with $|\Delta m|>1$, which are
not allowed in electric-dipole transitions. However, because of non-ideality
of the laser polarization state, instability of the magnetic field
and its direction, and Larmor precession, the $F'=1$ state of the
$^{2}S_{1/2}$ level becomes populated. An optical sideband ($\omega_{s}$
in Fig.~\ref{Flo:BeGeneral}) of the 313~nm wave used for laser
cooling, at 1.25~GHz offset, may be added with the purpose of depleting
this state and increase the laser cooling efficiency (repumper).

In this work, we used a two-photon, Doppler-free Raman transition
to precisely measure the Zeeman splitting and to determine the magnetic
field. In a $\Lambda$-scheme transition, two beams, pump and Stokes
($\omega_{c}$ and $\omega_{s})$, co-propagate in order to significantly
reduce the effect of Doppler broadening, by a factor $\simeq10{}^{6}$,
so that the residual Doppler linewidth is about $10^{2}$~Hz. 

A phase modulator inserted into the laser beam allows the simple generation
of the Stokes wave as a sideband; the fact of simultaneous carrier
and sideband generation is important for the stimulated Raman transition.
Thus, the light used for the Raman transition is the same as for the
purpose of laser cooling and repumping. The stimulated Raman transition
becomes strong for a case of the two-photon resonance, i.e. when $\Delta_{C}=\Delta_{S}$,
according to the notation in Fig~\ref{Flo:BeGeneral}. The latter
condition is strict for the case of $\Delta_{C},\,\Delta_{S}\gg\Omega_{C},\,\Omega_{S}$,
where $\Omega_{C}$ and $\Omega_{S}$ are the corresponding one-photon
Rabi frequencies. Generally, the two-photon transition is spectrally
broadened, which is characterized by the effective Rabi frequency
for a Raman transition. For our typical conditions, the sideband intensity
is varied in the range of a few percent to a few ten percent of the
carrier power, and also the total power is varied, and thus the effective
Rabi frequency is in the range of few kHz to many hundred kHz. The
a.c. Stark shift is of similar value. Large absolute sideband intensities
are conveniently used for preliminary determination of the Raman resonance.
Subsequently, a reduction of intensity is necessary for increasing
the measurement precision by reducing systematic shifts.

\section{Transitions and level splitting in $^{9}$Be$^{+}$}

Stimulated Raman transitions take place between Zeeman states of the
hyperfine levels $F=2$ and $F'=1$ of the $^{2}S_{\nicefrac{1}{2}}$
electronic ground state, with intermediate states being $F''=$0,
1, 2 or 3 of the electronically excited $^{2}P_{\nicefrac{3}{2}}$
level. The Zeeman splitting within one hyperfine state is (in the
following $F$ refers to any state) 
\[
E_{Z}=g_{F}m_{F}\mu_{B}\, B\simeq1.4\frac{\mathrm{MHz}}{\mathrm{G}}\, g_{F}m_{F}\, B\ ,
\]
where the $g_{F}$ factor is
\[
g_{F}\simeq\left(\frac{3}{2}+\frac{S(S+1)-L(L+1)}{2J(J+1)}\right)\frac{F(F+1)-I(I+1)+J(J+1)}{2F(F+1)}\ .
\]

The possible Zeeman components of the Raman transition $F'=1\rightarrow F=2$
and the corresponding Zeeman shifts (i.e. magnetic-field shift of
$\hbar(\omega_{c}-\omega_{s})$) are:
\[
\begin{array}{lcl}
\Delta E_{Z}=(\nicefrac{3}{2},\:1,\:\nicefrac{1}{2},\,0)\times1.4\frac{\mathrm{MHz}}{\mathrm{G}}\times B & \mathrm{for} & m_{F'}=-1\rightarrow m_{F}=(-2,\,-1,\,0,\,+1)\\
\Delta E_{Z}=(1,\:\nicefrac{1}{2},\:0,\:-\nicefrac{1}{2},\:-1)\times1.4\frac{\mathrm{MHz}}{\mathrm{G}}\times B & \mathrm{for} & m_{F'}=\phantom{-}0\rightarrow m_{F}=(-2,\,-1,\,0,\,+1,\,+2)\\
\Delta E_{Z}=(0,\:-\nicefrac{1}{2},\:-1,\:-\nicefrac{3}{2})\times1.4\frac{\mathrm{MHz}}{\mathrm{G}}\times B & \mathrm{for} & m_{F'}=+1\rightarrow m_{F}=(-1,\,0,\,+1,\,+2)
\end{array}
\]

for $\sigma^{-}$, $\pi$, and $\sigma^{+}$ transitions. Summarizing,
the possible Zeeman shifts of the Raman resonance at 1.25~GHz are:
\begin{equation}
\Delta E_{Z}=(0,\:\pm\nicefrac{1}{2},\:\pm1,\:\pm\nicefrac{3}{2})\times1.4\frac{\mathrm{MHz}}{\mathrm{G}}\times B\label{Total list-1}
\end{equation}
and a maximum of 7 different Zeeman shifts may be observed.

The excitation spectrum depends on strength of the magnetic field,
its direction with respect to the propagation direction of the cooling
laser beam and its sideband, on their polarization, and on their intensities.
However, we do not attempt to give a complete model, since it is not
necessary to do so for the present purpose.

It is important to note that a particular transition from the above
list Eq.~(\ref{Total list-1}) is observed only if the corresponding
initial $m_{F'}$-state of the $F'=1$ hyperfine level is populated.
Generally, all $m_{F'}$ states can be populated and thus all 7 lines
can be detected. There are special cases of $\sigma^{-}$ (and $\sigma^{+}$)-induced
transitions (carrier and sideband), for which only $m_{F'}=-1\ (+1)$
states of the $F'=1$ hyperfine level have significant population
in consequence of the interaction with the radiation. As mentioned
before, this is caused by Larmor precession, non-ideality of light
polarization state, and misalignment of magnetic field vector or field
strength fluctuation. The observed Zeeman splittings are then limited
to: 
\begin{equation}
\begin{array}{l}
\Delta E_{Z,\sigma^{+}}=-1.4\frac{\mathrm{MHz}}{\mathrm{G}}\times B\,,\;\mathrm{for}\:\sigma^{+}-\mathrm{induced\: transition\,,}\\
\Delta E_{Z,\sigma^{-}}=+1.4\frac{\mathrm{MHz}}{\mathrm{G}}\times B\,,\;\mathrm{for}\:\sigma^{-}-\mathrm{induced\: transition\,.}
\end{array}\label{Sigma +/- list-1}
\end{equation}

\section{Apparatus}

\begin{figure}
\begin{centering}
\includegraphics[width=11cm]{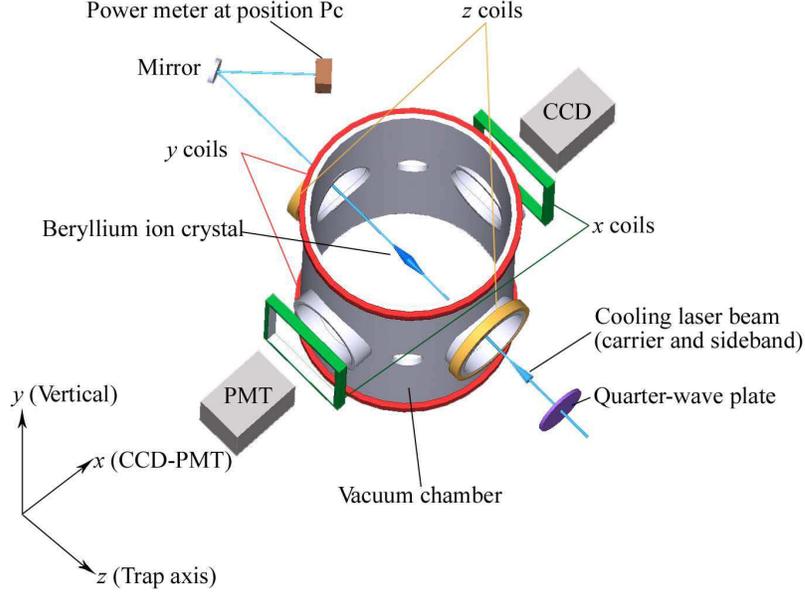}
\par\end{centering}

\protect\caption{Geometry of the ion trap apparatus and its magnetic field coils.}

\label{Flo:Setup}
\end{figure}

Fig.~\ref{Flo:Setup} illustrates the setup used for sympathetic
cooling and in which stimulated Raman spectroscopy of $^{9}$Be$^{+}$
ions has been studied. The cooling laser beam at 313~nm propagates
along the symmetry axis of the linear quadrupole trap ($z$-direction).
We denote the horizontal and vertical directions in the laboratory
as $x$ and $y$, respectively. The $^{9}$Be$^{+}$ fluorescence
is partially collected by a CCD camera and by a photomultiplier tube
(PMT) operating in the photon counting regime. The latter is used
to define the signal. The power of the cooling beam (carrier plus
sidebands) is measured after passing the vacuum chamber, at position
Pc.

For compensation of the external magnetic field, and for applying
a desired magnetic field, three pairs of magnetic field coils are
mounted around the vacuum chamber; the directions of the respective
fields are nearly orthogonal to each other.

The cooling laser is frequency-tunable and lockable in a detuning
range of a few hundred MHz relative to the $^{9}$Be$^{+}$ cooling
transition. Our laser (see Ref.~\cite{Vasilyev313nm}) includes internally
a waveguide electro-optic phase modulator driven at a frequency $f_{mod}\simeq1250\,$MHz
that generates a phase modulation sideband pair. The ``red'' sideband
is used as a repumper (wave $\omega_{s}$ in Fig.~\ref{Flo:BeGeneral}).

$^{9}$Be$^{+}$ ions are loaded and other species are removed from
the trap \cite{Blythe2005}. During this step, the 313~nm carrier
wave is red-detuned by a few hundred MHz from the cooling transition
line, and the sideband power is set to a few ten percent of the carrier
power. In order to cool the $^{9}$Be$^{+}$ ion ensemble and any
additional molecular ions to the lowest temperature, or to perform
measurements such as the ones described below, the carrier detuning
is instead kept at a few ten MHz, and the sideband power is typically
reduced by a moderate factor.

For performing stimulated Raman spectroscopy measurements, the sideband
frequency $f_{mod}$ is scanned, while all other parameters are kept
constant. 

\begin{figure}
\begin{centering}
\includegraphics[width=12cm]{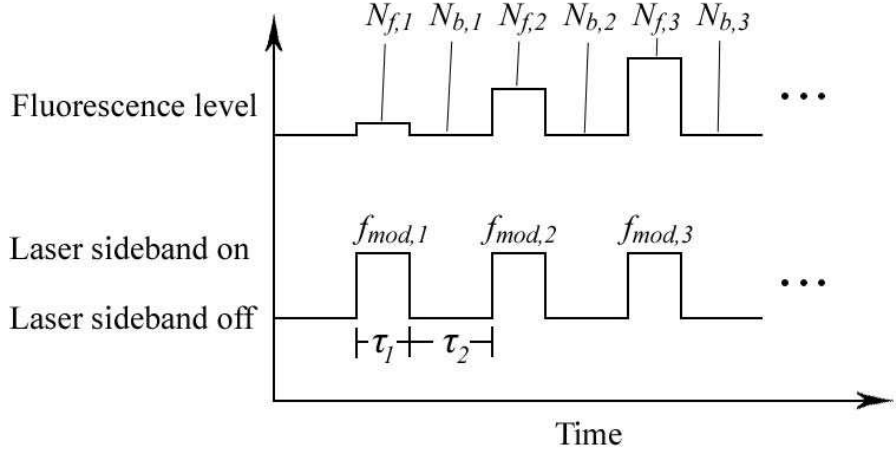}
\par\end{centering}

\protect\caption{Measurement time line. See the text for details. }

\label{Flo:DetScheme}
\end{figure}

The measurement sequence is illustrated in Fig.~\ref{Flo:DetScheme}.
During each cycle (duration $\tau_{1}+\tau_{2}$) the carrier is always
on, while the sideband is turned on and off. During the interval $\tau_{1}$
(sideband on) the PMT signal $N_{f,i}$ is recorded, and during the
interval $\tau_{2}$ (sideband off), the signal $N_{b,i}$ is recorded.
From one cycle to the next, the sideband frequency is changed from
$f_{mod,i}$ to $f_{mod,i+1}$, in order to scan the transition. For
each sequence $i$ of the frequency scan, we define the background-compensated,
normalized signal-to-noise ratio as the relative fluorescence $(N_{f,i}-N_{b,i})/N_{b,i}$.
This removes to some extent the influence of laser power variations,
assuming that the background count $N_{b}$ is proportional to laser
power (i.e. when it is due to scattered laser light).

In order to minimize the magnetic field in the trap region, an iterative
procedure was used. The magnetic field generated by one of the magnetic
field coil pairs was modified via its current and the current was
optimized so as to produce minimum Zeeman splittings, while the other
two field components were kept constant; then the value of the field
component of that first coil pair was held constant while the field
component generated by the second coil pair was varied, etc. Repeating
this procedure several times and also reducing cooling laser power
and sideband power in order to achieve higher resolution, we converged
to a small magnetic field strength.

\section{Experimental results and data analysis}

\begin{figure}
\begin{centering}
\includegraphics[width=11cm]{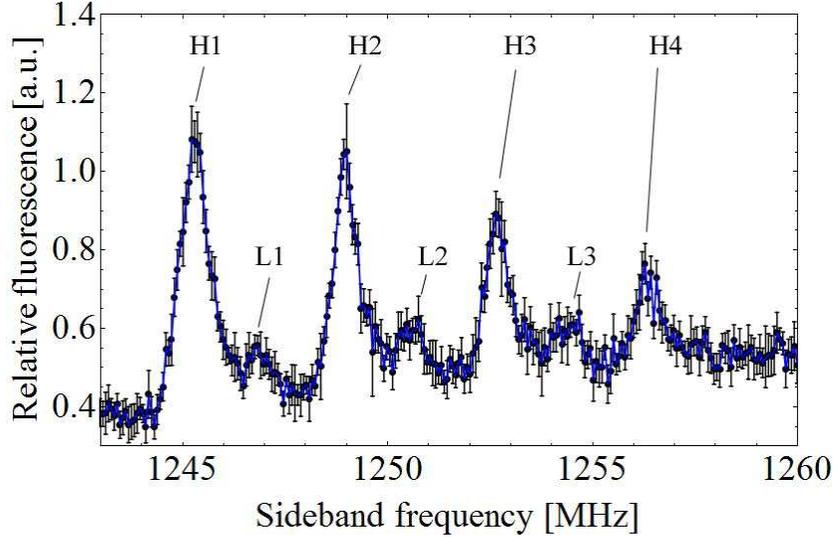}
\par\end{centering}

\protect\caption{Zeeman spectrum as a function of sideband frequency $f_{mod}$. All
7 possible transitions are detected and are denoted H1,..L3. Laser power
is 70~\textmu W. In this measurement, the current settings are $I_{x}=-5.74\textrm{ A, }$$I_{y}=1.70\textrm{ A, }$$I_{z}=0.14\textrm{ A}$,
which correspond to $B_{x}=-2.434\textrm{\,\ G, }$$B_{y}=0.008\textrm{\,\ G, }$$B_{z}=1.022\textrm{\,\ G}$.}

\label{Flo:7peaks}
\end{figure}

\begin{figure}
\begin{centering}
\includegraphics[width=11.9cm]{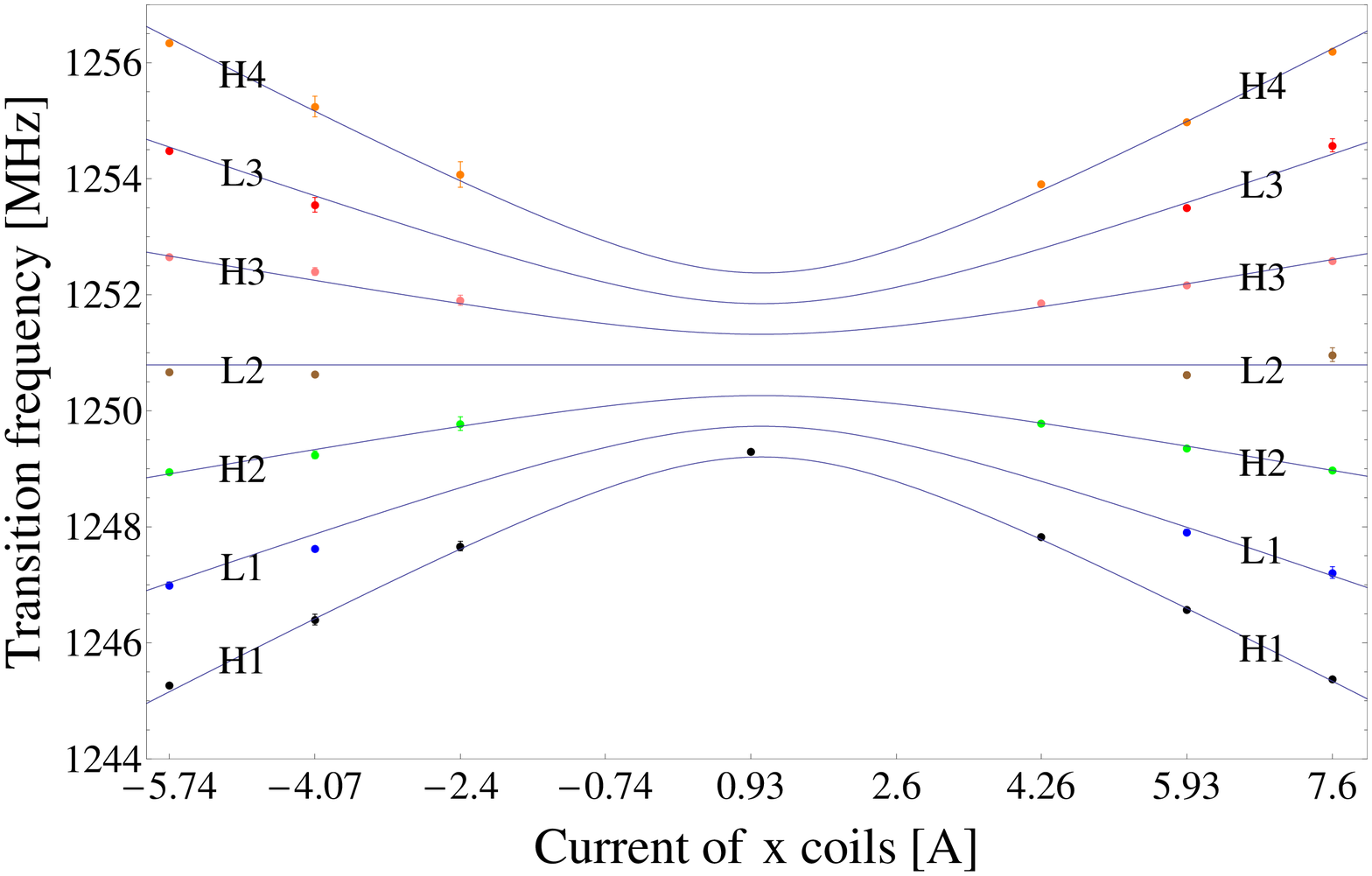}
\par\end{centering}

\begin{centering}
\includegraphics[width=11.9cm]{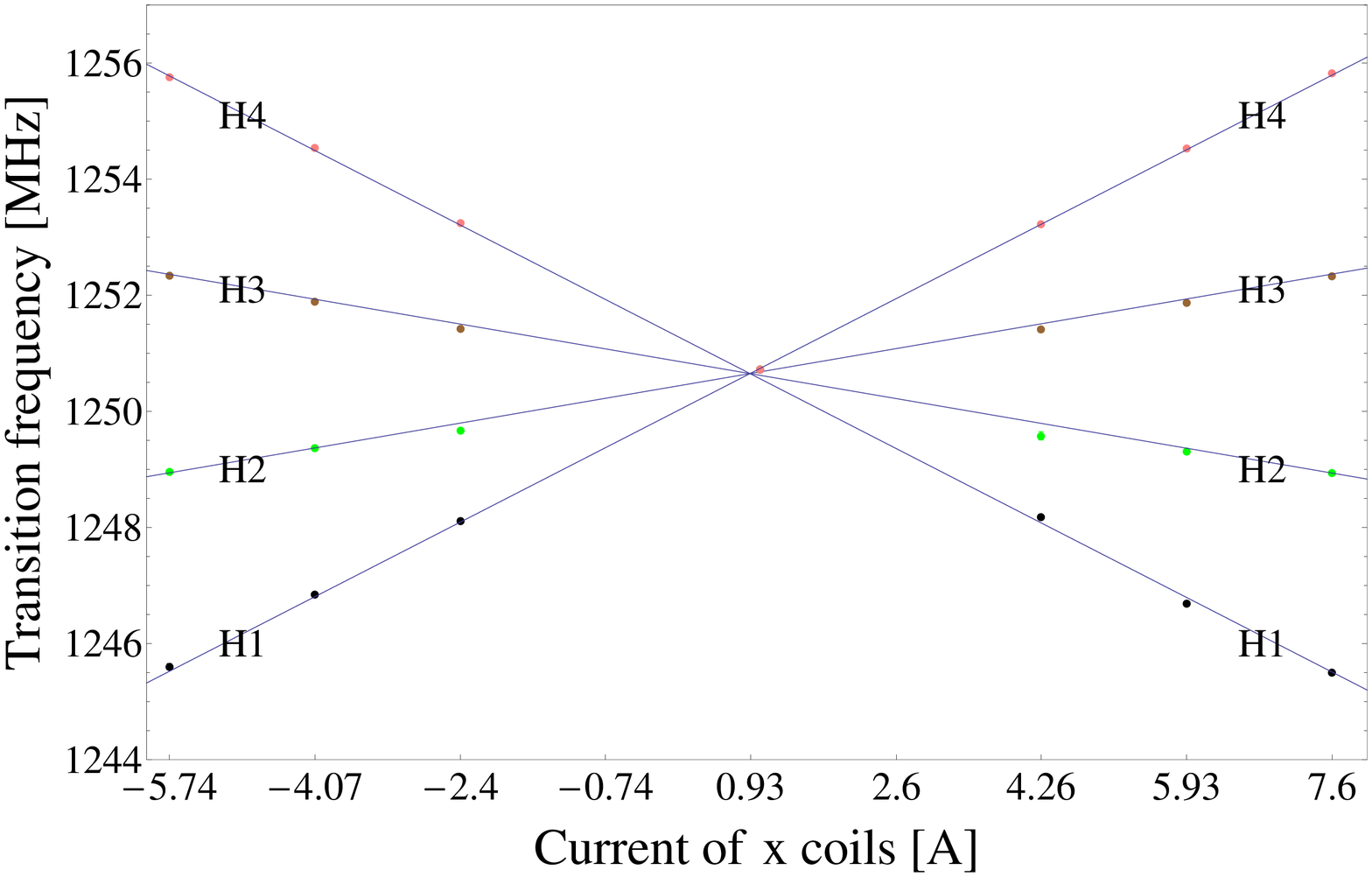}
\par\end{centering}

\protect\caption{Zeeman shifts as a function of one magnetic field component, $B_{x}$.
Top: Data set I. The data points are the transition frequencies obtained
from individual scans such as Fig.~\ref{Flo:7peaks}, all at 70~\textmu W
power. The frequencies of that figure are reported in the present
plot as data points for the current $I_{x}=-5.74$~A. $B_{y}$ and
$B_{z}$ are not zero: $B_{y}=0.008\textrm{ G, }$$B_{z}=1.022\textrm{ G}$
($I_{y}=1.70\textrm{ A, }$$I_{z}=0.14\textrm{ A}$). Bottom: $B_{y}$
and $B_{z}$ are set to close to zero ($I_{y}=1.681\textrm{ A, }$$I_{z}=-0.166\textrm{ A}$). }

\label{Flo:FieldScan}
\end{figure}

In the general case, when all transitions take place, Eq.~(\ref{Total list-1})
applies. Fig.~\ref{Flo:7peaks} illustrates the Zeeman spectrum for
a particular, fixed applied magnetic field. Fig.~\ref{Flo:FieldScan}~(top)
shows the Zeeman splittings for several different values of one magnetic
field component, here the $B_{x}$-component. In this measurement
it can be directly seen that the transverse magnetic field $B_{y}^{2}+B_{z}^{2}$
is nonzero. For each measurement, the cooling light power was recorded
as well. 

We also performed measurements similar to the one shown in Fig.~\ref{Flo:FieldScan}
but varying the magnetic field component, $B_{y}$ (data set II) and
$B_{z}$ (data set III). The data sets I, II, III, all taken at the
same power (70~$\mu$W at position Pc), were fitted with the functions:
\begin{equation}
\begin{array}{l}
f_{\eta}=\eta\,\sqrt{B_{x}^{2}+B_{y}^{2}+B_{z}^{2}}+f_{offset}\,,\\
B_{j}=k_{j}\, I_{j}+B_{j,offset}\,,
\end{array}\label{eq:Fit formula-1}
\end{equation}
where $j$ denotes the spatial components $x,\, y$ and $z$. The
values of $\eta$ are taken in accordance with the list Eq.~(\ref{Total list-1}).
The seven coefficients $B_{j,offset}$, $k_{j}$, and $f_{offset}$
were obtained from a least-squares fit. The result of this fit may
be expressed as:

\begin{equation}
\begin{array}{ll}
B_{x}= & (0.362\pm0.003)\textrm{ G/A}\times[I_{x}-(0.985\pm0.042)\mathrm{\textrm{ A}}]\,,\\
B_{y}= & (0.434\pm0.049)\mathrm{\textrm{ G/A}}\times[I_{y}-(1.681\pm0.065)\mathrm{\textrm{ A}}]\,,\\
B_{z}= & (3.586\pm0.036)\mathrm{\textrm{ G/A}}\times[I_{z}+(0.145\pm0.007)\mathrm{\textrm{ A}}]\,.
\end{array}\label{eq:global fitting}
\end{equation}

From this fit, we obtain the current values that minimize the magnetic
field components. After setting the components $B_{y}$ and $B_{z}$
close to zero, we recorded data again as a function of $B_{x}$. The
result is shown in Fig.~\ref{Flo:FieldScan}~(bottom). The effect
of the minimization of $B_{y}$ and $B_{z}$ is obvious. Note that
in this case, the peaks L1, L2 and L3 are not detected any more. 

\begin{figure}
\begin{centering}
\includegraphics[width=12cm]{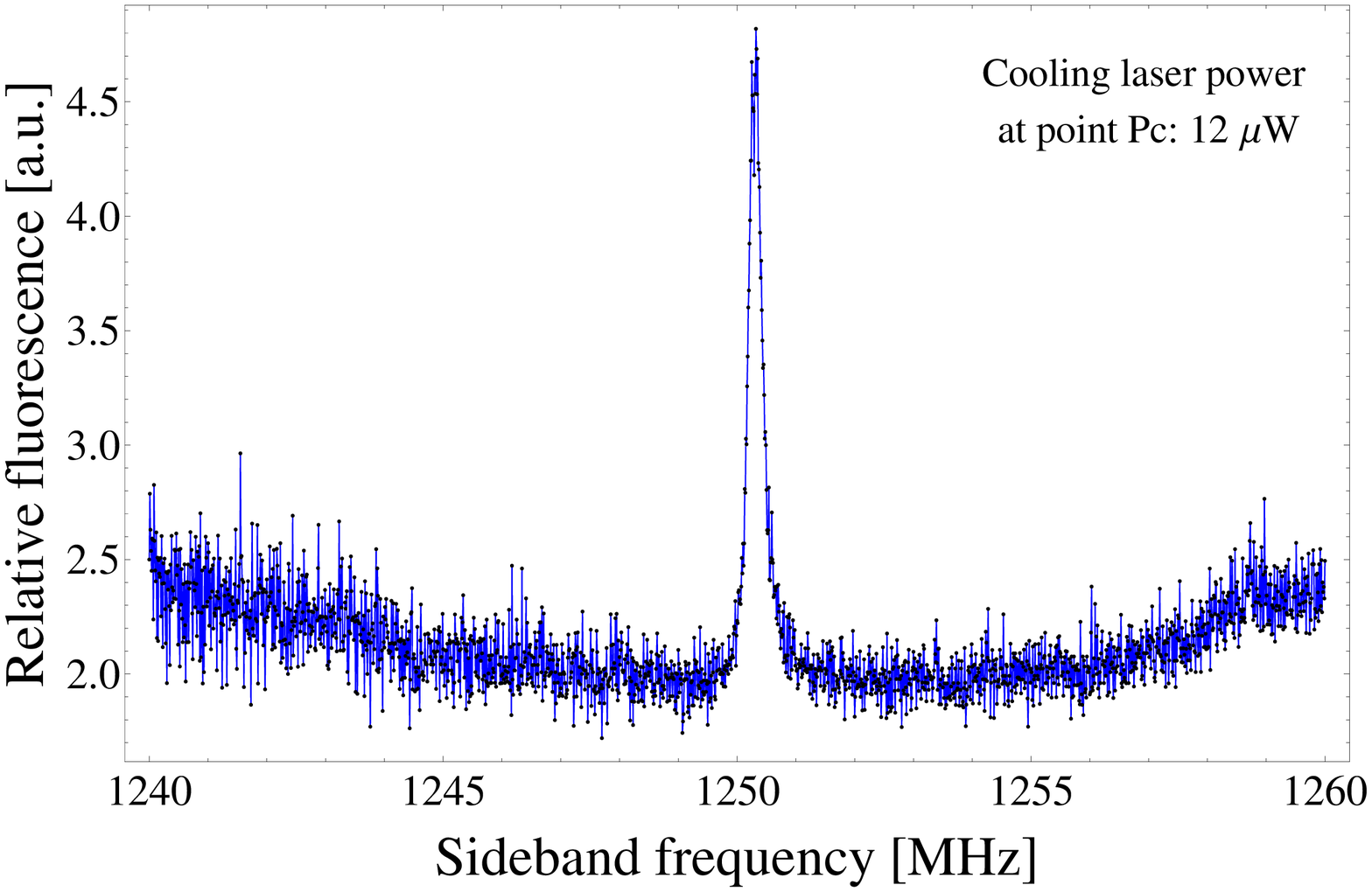}
\par\end{centering}

\protect\caption{Spectrum obtained with a $\sigma$-polarized beam. Only one peak is
detected. A nonzero magnetic field in the beam propagation direction,
$B_{z}\protect\ne0$, is applied. The other field components are nearly
zero.}

\label{Flo:SinglePeak}
\end{figure}

\begin{figure}
\begin{centering}
\includegraphics[scale=0.38]{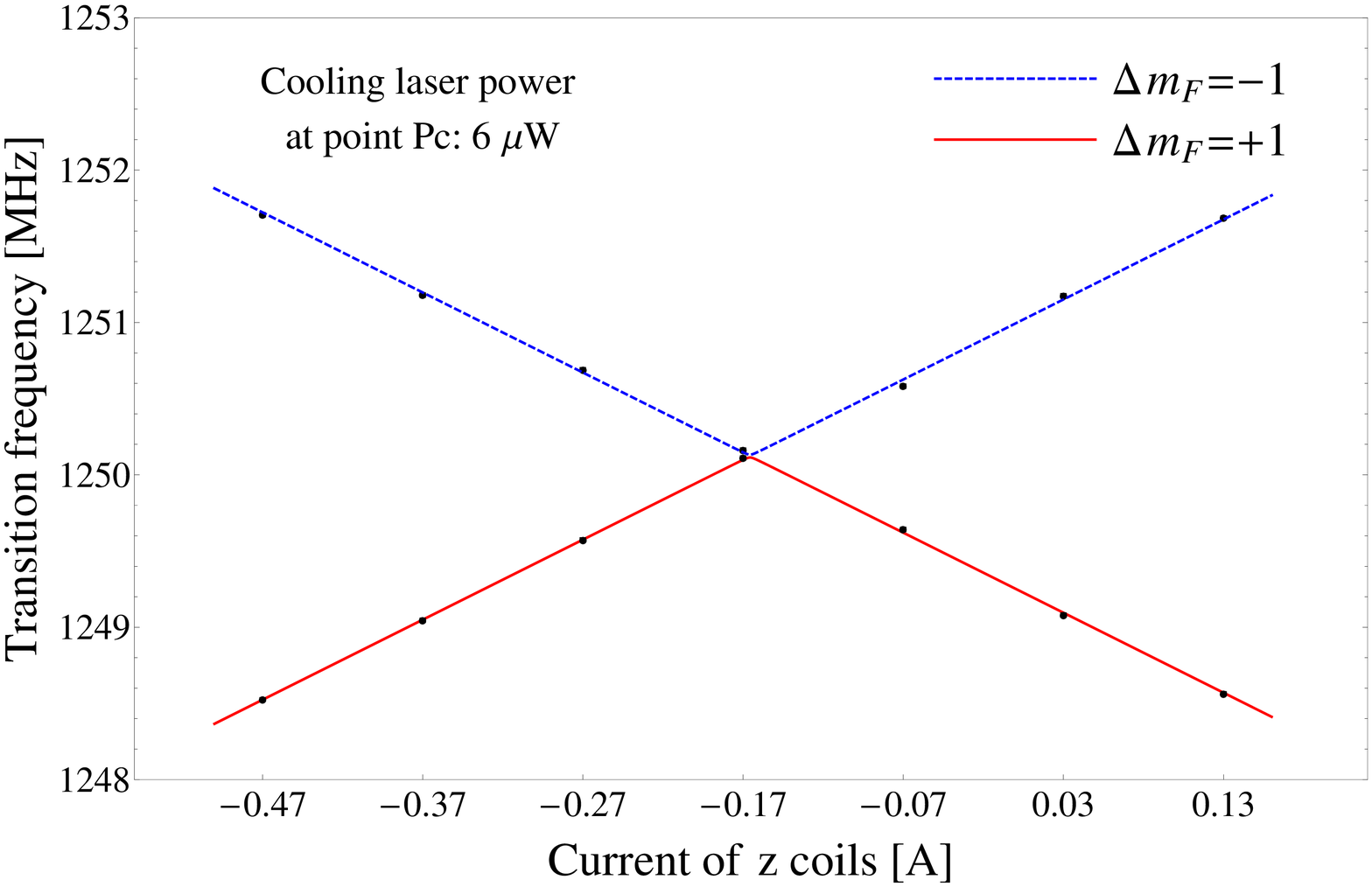}
\par\end{centering}

\protect\caption{\label{fig:Magnetic-field-scan along z} Zeeman shift of the $\Delta m_{F}=m_{F'}-m_{F}=\mp1$
transitions as a function of $B_{z}$. The magnetic field components
$B_{x}$ and $B_{y}$ are minimized.}
\end{figure}

Subsequently, we set $B_{x}$ and $B_{y}$ close to zero, and varied
$B_{z}$. Now the laser wave can be set to $\sigma$ polarization
since the magnetic field is then along the direction of propagation.
A typical spectrum under a fixed value of $B_{z}$ is shown in Fig.~\ref{Flo:SinglePeak}.
There is only a single resonance. Varying now $B_{z},$ we obtain
data set IV, which is displayed in Fig.~\ref{fig:Magnetic-field-scan along z}.
It was taken with a low value of beam power, 6~$\mu$W at position
Pc. 

\begin{figure}
\begin{centering}
\includegraphics[width=11cm]{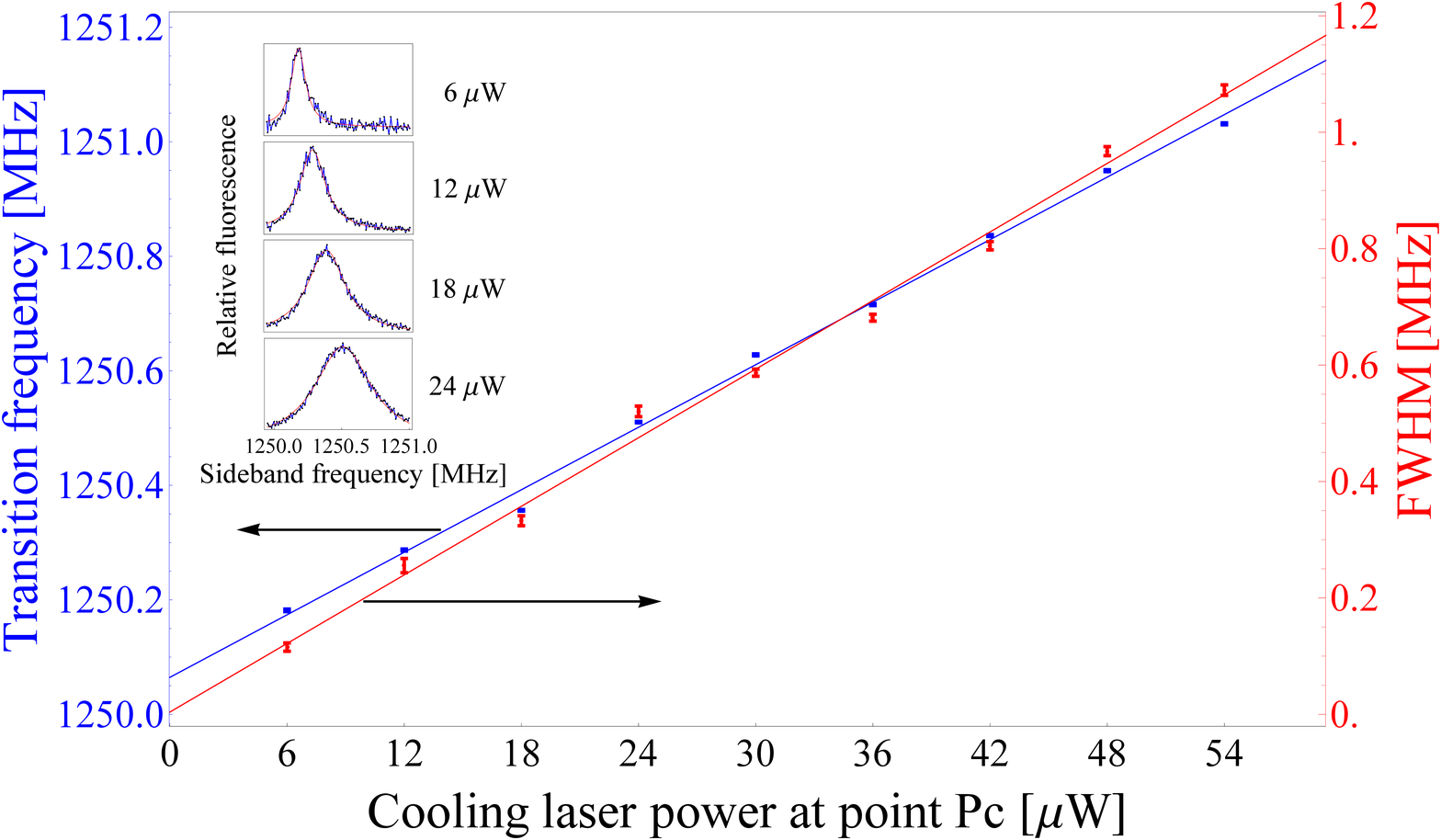}
\par\end{centering}

\protect\caption{Light shift and power broadening as a function of laser power. Inset:
light shift and power broadening of a $\sigma$-transition. All spectra
were taken under the same conditions, only the beam power was varied.
The magnetic field strength is nearly zero. }

\label{Flo:LineShiftAndBroadening}
\end{figure}

We also investigated the light-atom interaction effects. The peak
shift and the linewidth as a function of beam power are summarized
in Fig.~\ref{Flo:LineShiftAndBroadening}. The inset of Fig.~\ref{Flo:LineShiftAndBroadening}
shows examples of the power dependence of the peak frequency $f_{\eta}$
and of the line shape of one of the two $\sigma$-transitions under
near-zero field conditions. We find a linear dependence of the frequency
shift and linewidth on the light power. Extrapolation of the peak
frequency to zero beam power predicts the hyperfine frequency to be
$f_{0}=1250.065\pm0.013$~MHz. A value of $1250\,017\,670.46\pm1.5$~Hz
has been obtained by Nakamura et al \cite{NakamuraBe+RF}, and a value
of $1250\,017\,674.088\pm0.024$~Hz has been obtained by Shiga et
al \cite{ShigaItanoBollingerBe+2011}; the discrepancy of our measurements
from these values ($3.6\:\sigma$) may in part be due to an imperfect
polarization state of the beam or the small residual magnetic field.

\section{Discussion}

In this work, we focus on the ability to minimize the magnetic field.
We consider this issue from two points of view. First, the fit of
data sets I, II, III allows us to compute the magnetic field uncertainty
when the currents are set to minimize the field components. The result
is 

\[
\Delta B=\sqrt{(0.362\times0.042)^{2}+(0.434\times0.065)^{2}+(3.586\times0.007)^{2}}{\rm \, G}=0.041\,{\rm G}\:.
\]
We can also consider the data set IV (Fig.~\ref{fig:Magnetic-field-scan along z}).
A fit of only this data yields

\begin{equation}
\begin{array}{rcl}
B_{z} & = & (3.753\pm0.020)\mathrm{\textrm{ G/A}}\times[I_{z}+(0.166\pm0.001)\mathrm{\textrm{ A}}]\\
\sqrt{B_{x}^{2}+B_{y}^{2}} & = & (0.007\pm0.031)\textrm{ G\,.}
\end{array}\label{eq:fit z direction}
\end{equation}

The result for the current value that minimizes $B_{z}$ lies outside
the $2\sigma$ uncertainty range of the result obtained in Eq.(\ref{eq:global fitting}),
but it should be noted that the conditions of data set IV were different:
lower beam power and smaller range of offsets from the zero-field-component
current value. Nevertheless, the field strength uncertainty, $\Delta B'=\sqrt{(3.753\times0.001)^{2}+(0.031)^{2}}{\rm \, G}=0.031\,{\rm G}\:$
is comparable to $\Delta B$.

As a third result, we can consider the measurement performed with
near-zero field, and consider the difference between the zero-intensity
extrapolated resonance frequency of Fig.~\ref{Flo:LineShiftAndBroadening}
and the literature experimental value, $\Delta f_{0}=47\pm13\,{\rm kHz}$.
From this we may set an upper limit to the field strength present
during that measurement, $(47+13)\,{\rm kHz/(1.4\,{\rm MHz/G})}\simeq43\,$mG.

Finally, we can provide an upper limit for the magnetic field gradient
as follows. The narrowest linewidth observed when reducing the laser
power as much as possible while keeping the ions cooled (approximately
2~$\mu$W), is approximately 40~kHz. Assigning the linewidth to
be due completely to a nonzero magnetic field gradient across the
ion ensemble (size: 2~mm in $z$-direction), it would then be equal
or less to 40~kHz/(1.4~MHz/G) $\simeq29$~mG. This is a conservative
upper limit, since Fig.~\ref{Flo:LineShiftAndBroadening} indicates
that at this laser power there is still light-induced broadening present.

\section{Conclusion and outlook}

We demonstrated a simple method for measuring the magnetic field in
an ion trap that uses beryllium ions. The method is simple since it
utilizes only the cooling laser and an optical sideband that is also
conventionally present since it serves as a repumper during laser
cooling. The magnetic field resolution, approximately 40~mG, is at
present limited by power broadening and the Stark shift induced by
the cooling wave, imperfect polarization state of the cooling wave,
and by the temporal instability of the currents supplied to the magnetic
field coils (here, the instability is of the order of 1~mA). In addition,
a non-uniformity of the magnetic field, if present, also limits the
resolution of the absolute field strength determination. 

The method has been applied here to an ensemble containing a large
number of trapped ions (a few thousands), since this is typical for
our application of sympathetic cooling. We believe that the method
could be applied also in the case of few ions or a single trapped
ion. In the case of a moderate number of trapped ions (e.g. an ion
string), the use of a tightly focused beam would allow addressing
an individual ion and determining the field at its location. 

As a further perspective, we propose that it may be possible to eliminate
the light shift. This requires equal Rabi frequencies of the sideband
and the carrier; then the light shifts induced by the two radiation
fields compensate each other. The one-photon transition induced by
the strong sideband, can, in principle, hinder from observation the
two-photon transition, but an appropriate choice of intensities and
detunings can allow for (partial) compensation of the light shifts
and for detection of the two-photon signal. A complete compensation
of the frequency shifts means equality of sideband and carrier intensities
(neglecting for simplicity the issue of branching ratios). Thus, this
scheme requires a modulation index of about 0.33, which is available
from high-efficiency waveguide modulators.

\medskip{}
Acknowledgement: We are indebted to Q.-F. Chen for assistance in the
data analysis. We are also grateful to M. Hansen and U. Bressel
for performing a magnetic field characterization using magnetic sensors.
This work was done in the framework of project Schi 431/19-1 funded
by the Deutsche Forschungsgemeinschaft.

\bibliographystyle{apsrev4-1}
\bibliography{mylibrary}

\end{document}